\documentstyle[preprint,prl,aps]{revtex} 
\begin{document}
\draft
\title{Linear Temperature Variation of the Penetration Depth in
$\bf YBa_{2}Cu_{3}O_{7\bbox{-\delta}}$ Thin Films}
\author{L.A.\ de\ Vaulchier, J.P.\ Vieren, Y.\ Guldner, N.\ Bontemps}
\address{Laboratoire de Physique de la Mati\`ere Condens\'ee, Ecole Normale
Sup\'erieure, 24 rue Lhomond, 75231 Paris Cedex 05, France}
\author{R.\ Combescot}
\address{Laboratoire de Physique Statistique, Ecole Normale Sup\'erieure,
24 rue Lhomond, 75231 Paris Cedex 05, France}
\author{Y.\ Lema\^\i tre and J.C.\ Mage}
\address{Laboratoire Central de Recherches, Thomson-CSF,Domaine de
Corbeville, 91404 Orsay Cedex, France}
\date{Received \today}
\maketitle
\begin{abstract}
We have measured the penetration depth $\lambda(T)$ on
$\rm YBa_{2}Cu_{3}O_{7}$ thin films from transmission at
120, 330 and 510~GHz, between 5 and 50~K. Our
data yield simultaneously the absolute value and the temperature
dependence of $\lambda(T)$. In high quality films $\lambda(T)$ exhibits the
same linear temperature dependence as single crystals, showing its
intrinsic nature, and $\lambda(0)=1750\,{\rm \AA}$.
In a lower quality one, the more usual $T^2$ dependence is found, and
$\lambda(0)=3600\,{\rm \AA}$. This suggests that the $T^2$ variation
is of extrinsic origin. Our results put the
$d$-wave like interpretation in a much better position.\\
\end{abstract}
\narrowtext
The puzzling properties of high-$T_c$ superconductors have stimulated
recently extensive experimental and theoretical work in order to determine
the actual symmetry of the order parameter. While in the BCS weak coupling
theory, the penetration depth $\lambda(T)$ varies as
$exp(-{\it\Delta}/{k_B T})$,
a quadratic behavior $\lambda(T)\sim T^2$ was reported by many groups in YBCO
thin films \cite{annett,porch,ma}. Recently, in very high quality YBCO single
crystals, a linear temperature dependence was measured up to 40~K
\cite{hardy,zhang,mao}. Such a variation, never observed so far
in YBCO thin films \cite{beasley}, is consistent with the occurrence of nodes
in the gap and may suggest a $d$-wave pairing mechanism
\cite{pines}. However other possible interpretations have been raised, both
theoretically and experimentally, e.g. a possible proximity effect with a
normal metal layer \cite{mao,pambianchi}, or the sensitivity of a
conventional BCS type $\lambda(T)$ to the oxygenation of the samples
\cite{kresin,xue}. Several reasons for the discrepancy observed between thin
films and single crystals may be invoked: (i) within the framework of
$d$-wave pairing, scattering due to impurities or defects may change the
linear temperature dependence into a quadratic one in thin films
\cite{sauls}, where such defects would be more numerous. However $T_c$
should then also be affected, which is not the case \cite{hardy,prohammer}.
(ii) weak links, more likely to be present in thin films, may yield  an
effective penetration depth (larger than the intrinsic one)
with a different temperature dependence \cite{hylton,halbritter}.
The penetration depth $\lambda$
has been essentially measured in both single crystals and thin films by
using surface impedance techniques at a single frequency
\cite{porch,hardy}. Such measurements actually measure the
variation $\Delta\lambda(T)=\lambda(T)-\lambda(0)$, but without the
knowledge of
$\lambda(0)$ on each sample, comparing the temperature dependence of
various samples may not be significant. This paper reports for the first
time, from high frequency transmission data, both an unambiguous linear
temperature dependence of $\lambda(T)$ on high quality YBCO thin films,
which is quantitatively the same as in single crystals \cite{hardy},
and  simultaneously $\lambda(0) \sim 1750\,{\rm\AA}$. We observe a
quadratic temperature dependence on a lower quality sample, and
$\lambda(0) \sim 3600\,{\rm\AA}$. We
suggest that this $T^2$ dependence in thin films is not a consequence of
strong scattering (in contrast with Zn or Ni doped materials
\cite{bonn2}) but is of extrinsic origin. The intrinsic dependence of
$\lambda(T)$ at low temperature is linear. This
linear dependence points plainly toward the existence
of low energy excitations, but our results alone do not allow to identify
them. A natural and popular explanation is that they are due to nodes
in the gap arising from unconventional pairing of $d$-wave type
\cite{annett,hardy,zhang}, an interpretation also conveyed by
SQUID experiments \cite{wollman,tsuei}. Our results provide support to
$d$-wave pairing, because they show a linear behavior in films, but
also because they eliminate the uncomfortable need to explain the $T^2$
dependence by impurity scattering at the unitary limit \cite{sauls}.\\

The experimental transmission set-up uses carcinotron tubes as powerful,
stable microwave sources and oversized waveguides in order to change easily
the frequency  (120 to 510~GHz) \cite{LApc,LAprb}. We took an extreme care
to lower microwave leakages down to 60~dB. The transmitted signal is
detected by a helium cooled InSb bolometer. Measurements have
been performed by slowly varying the temperature between 5 and 110~K at fixed
frequencies. The transmission of
the substrate was checked to vary by less than 1~\% between 5 and
110~K. We have screened the samples from stray magnetic field, so
that the residual field is less than 0.5~Gauss.
For a film of thickness $d$ (smaller than the skin depth in the normal
state or the penetration depth in the superconducting state) deposited onto a
substrate of index $n$, the transmission writes \cite{glover}:
\begin{eqnarray}
{\rm T} = {1\over \left| 1+{\sigma d Z_0 \over 1+n} \right|^2}
\label{eq1}
\end{eqnarray}
where $\sigma=\sigma_1-i\sigma_2$ is the complex conductivity of the film,
$Z_0$ the impedance of free space and $Z=1/\sigma d$ the impedance of the
film. If in the energy and temperature range of
interest $\sigma_2(T) \gg \sigma_1(T)$ and
${\sigma_2 d Z_0 \over 1+n} \gg 1$, we can write:
\begin{eqnarray}
{\rm T \over T_{110}}=\mu^2_0\, \omega^2 \lambda^4(T) \, \sigma^2_{110}
\label{eq2}
\end{eqnarray}
where $\sigma_{110}$ is the normal 110~K conductivity of the film. We choose
to normalize the transmission at 110~K because the ratio thus obtained is
experimentally more reliable than the absolute value of the transmission.
For simplicity we assumed  ${\sigma_{110} d Z_0 \over 1+n} \gg 1$ in
(\ref{eq2}). Thus, the measurement of T/T$_{110}$
yields an absolute value of $\lambda(T)$. The uncertainties on the
$\lambda(T)$ value which arise when neglecting the interferences within the
film and/or the substrate, and the effect of the finite film thickness
with respect to $\lambda(T)$ have been estimated: the transmission is
fairly insensitive to the interference effects at our frequencies as can be
shown by comparing a complete expression for T to (\ref{eq1}). Moreover, no
significant change of the transmitted energy could be observed in the most
sensitive range 440--550~GHz \cite{LAprb}. The finite film thickness,
yields an approximate
80~\AA\ overestimate of $\lambda(T)$ for $d/\lambda(T) \sim 0.5$.
$\sigma_2(\omega,T)\gg\sigma_1(\omega,T)$ holds at low
temperature $(T\le 20~{\rm K})$ for all frequencies, but should break down at
40~K \cite{zhang,bonn,buhleier}. Indeed,
$\sigma_1(35\,{\rm GHz, 40\,K}) \sim 2\times 10^7\,\Omega^{-1} {\rm m}^{-1}$,
a value likely larger than $\sigma_2(300\,{\rm GHz})$. However we expect a
strong decrease of $\sigma_1(T)$ at 300~GHz \cite{bonn,buhleier,dahne}.\\

It is essential that we define the criteria we use in order to sort out
the films that we investigate. Our films are epitaxially grown either by
laser ablation on MgO \cite{gv} or by sputtering on LaAlO$_3$
\cite{lemaitre}. They have a narrow transition $(\Delta T_c \le 1\,{\rm K})$.
$T_c$ is 86--92~K, depending on the substrate. In previous papers, the
correlation between the film quality in terms of transmission, width of the
rocking curve and surface resistance has been established
\cite{LApc,LAprb,thivet}. This led us to select the films either
from their surface resistance: $R_S \le 0.5\,{\rm m} \Omega$ at 77~K and
10~GHz, or the width of their rocking curve
$\Delta \theta \le 0.5^{\circ}$. The characteristics of our samples
are listed in Table~\ref{table1}. Two types of films have
been intentionally investigated. The first film ($A1$) is of poor
quality with respect to the above criteria. The two others ($B1$ and
$B2$) display either a low surface resistance or a narrow rocking curve. We
show in Fig.~\ref{L439Ytrans} and Fig.~\ref{HC65trans} the normalized
transmission of the $A1$ and $B2$ films respectively. Strong differences
are observed, in particular for  the residual transmission in the
superconducting state. At $T \le 0.5\,T_c$, the transmission increases as
$\omega^2$ in sample $B2$, as expected from (\ref{eq2}) and as shown in the
inset of Fig.~\ref{HC65trans}. In sample $A1$,
the data can be analyzed by adding a constant T$_0$ to the $\omega^2$ term
\cite{footnote}. We estimate T$_0$ from the 10~K data (see inset of
Fig.~\ref{L439Ytrans}). We compute $\lambda(T)$ from (\ref{eq2}) for the
three frequencies, assuming that T$_0$ in the case of $A1$, does not depend
on temperature or frequency. The result of this analysis is
shown on Fig.~\ref{L439Ylambda} and Fig.~\ref{HC65lambda}, for the
samples $A1$ and $B2$. The 120, 330 and 510~GHz curves
collapse, up to 40~K for sample $A1$ and up to 55~K for sample $B2$. This
confirms that the frequency dependent part of the transmission varies as
$\omega^2$, and that $\sigma_2(\omega,T) \gg \sigma_1(\omega,T)$ in these
temperature ranges. $\lambda(0)$ is found
to be $1750 \pm 160 \,{\rm\AA}$ for $B2$ and $3600 \pm 200 \,{\rm\AA}$ for
$A1$. The uncertainty on $\lambda(0)$ depends mostly on the accuracy on the
$\sigma_{110}$  measurement, which is limited by the uncertainty on the film
thickness $(\pm 100\,{\rm\AA})$ and by the Van der Pauw technique.
Finally the most striking result is the temperature dependence of
$\lambda(T)$. For sample $A1$, we find as shown in the inset, a clear $T^2$
dependence, a fairly common result for thin films \cite{annett,porch,ma},
associated with a very large value for $\lambda(0)$. In
contrast, for sample $B2$, we find a linear temperature dependence up to
50~K along with a much shorter $\lambda(0)$. Such a linear behavior was
similarly observed in film $B1$.\\

We now discuss the implications of these results. We believe that they
demonstrate clearly the extrinsic origin of the $T^2$ dependence found
previously in thin films. Indeed they agree with a phenomenological
expression $\lambda^2(T)=\lambda^2_{intr}(T)+\lambda^2_{extr}(T)$ as proposed
by Hylton {\it et al.}, with $\lambda_{extr}$ being for example the
contribution of weak links (but our conclusions are obviously independent of
this specific expression). When $\lambda_{extr}$ is the dominating
contribution as in our film $A1$ it may produce the $T^2$ dependence which is
likely to change from sample to sample. Indeed, the data of
Porch {\it et al.} (see inset of Fig.~\ref{L439Ylambda}) are not the same as
ours. On the other hand, for good enough films, $\lambda_{extr}$ gets
negligible and we find the intrinsic behavior for $\lambda(T)$ which is
linear, and is expected to be unchanged from films to crystals. The YBCO
single crystals results of Hardy {\it et al.} are reported on
Fig.~\ref{HC65lambda}, shifted to our $\lambda(0)$ value. The excellent
agreement between the two sets of data confirms
the intrinsic nature of $\lambda$ in film $B2$.
We remark that the agreement between $\lambda(T)$ in two
completely unrelated samples is a very strong result since, except for a
coincidence, it eliminates any extrinsic explanation like poor surface
effects or pair breaking due to magnetic impurities \cite{kresin}, for the
interpretations of the linear dependence. However the nature of the low
energy excitations is still unknown \cite{neminsky,eliashberg} and it is also
not yet clear whether they are linked to the planes or to the chains.
Furthermore, the $T^2$ dependence appears in a film where the penetration
depth is large. Our $\lambda_{intr}(0)$ is much shorter. It is larger
though than $\lambda_{ab}(0)=1450$--1490~\AA\ derive by $\mu$SR
\cite{sonier}, and consistent with $\lambda_a(0)=1600$~\AA\ provided by
infrared reflectance data \cite{basov}. We believe that most techniques
do not provide as a straightforward determination as ours.\\

Our results offer a solution to a previous uncomfortable situation
arising in the $d$-wave interpretation of the experimental results for
$\lambda(T)$. Indeed in order to explain the $T^2$ behavior in films
together with the linear dependence in single crystals, theoretical
calculations had to call for a high impurity concentration in films. But this
implies a significant difference between $T_c$ for films and crystals which
is not observed. The proposed escape \cite{hirschfeld} was to assume that the
impurity concentration was rather low with a scattering very near the
unitary limit, a very specific hypothesis. In this case $T_c$ could be
little changed by impurities while the low $T$ behavior of $\lambda(T)$
would be much more affected. Since our results show that the $T^2$ behavior
is extrinsic, e.g. due to weak links \cite{LAprb}, all the results become
coherent with a reasonably low content of ordinary scatterers.\\

In order to interpret quantitatively our results, it is convenient to make
use of the simplified two-dimensional model of $d$-wave like pairing
introduced by Xu {\it et al.} \cite{sauls}, where the gap has a linear
dependence near the nodes and is constant elsewhere:
$\left|{\it\Delta}(\theta)\right|=\mu {\it\Delta}_0 \theta$ for
$0\le\theta\le 1/\mu$ and $\left|{\it\Delta}(\theta)\right|={\it\Delta}_0$
for $1/\mu \le\theta\le\pi/4, \theta$ being the angle with respect to the
node position on the Fermi surface and the rest of ${\it\Delta}(\theta)$
being obtained by symmetry. The density of states is taken as constant. This
model contains the essential physical ingredients to represent a general
order parameter with $d$-wave symmetry, except for the possible effect of
Van Hove singularities. The slope of $\lambda(T)$ at $T=0$ is given by
$2\,ln\,2 / \mu {\it\Delta}_0$, however
when ${\it\Delta}_0$ is calculated with a weak coupling assumption, the
result can only be made to agree with experiment up to 20~K and it does not
reproduce the remarkable linear behavior up to 55~K.
A simple way to account for this feature is to incorporate strong coupling
effects. This is justified on theoretical ground \cite{pines} and is also
consistent with experiments since for example tunneling data give an
effective gap larger than what is expected from BCS theory. If we take
$2 {\it\Delta}_0 / T_c = 7$ (a typical tunneling value), we obtain the
solid curve on Fig.~\ref{HC65lambda} which agrees quite well with our
experimental results. This gives further support to this interpretation. In
summary, although our results cannot be taken as a proof for $d$-wave
pairing, they put at least this interpretation in a much better position.\\

We are deeply indebted to M.\ Guilloux-Viry, C.~Le~Paven-Thivet and
A.\ Perrin from the Laboratoire de Chimie Inorganique et Mol\'eculaire,
Universit\'e de Rennes, for providing us with well characterized thin
films. We thank P.\ Monod for his critical reading of the manuscript.
This work has been supported by EEC Contract 93--2027.IL and by the
New Energy and Industrial Technology Development Organization (NEDO). Both
laboratories Physique de la Mati\`ere Condens\'ee and Physique Statistique
are associated to CNRS and Universities Paris~VI and Paris~VII.
%

\begin{figure}
\caption{110~K normalized transmission versus temperature for the YBCO $A1$
film. The inset shows the $\omega^2$ law of the normalized transmission at
10~K.
\label{L439Ytrans}}
\end{figure}
\begin{figure}
\caption{110~K normalized transmission versus temperature for the YBCO $B2$
film. The inset shows the $\omega^2$ law of the normalized transmission
at 10~K.
\label{HC65trans}}
\end{figure}
\begin{figure}
\caption{Temperature dependence of $\lambda(T)$ for the YBCO $A1$ film. The
$T^2$ law (see inset) is characteristic of a poor quality film. The bold
squares represent the 8~GHz Porch {\it et al.} \protect \cite{porch} data
points shifted to our $\lambda(0)$ value.
\label{L439Ylambda}}
\end{figure}
\begin{figure}
\caption{Temperature dependence of $\lambda(T)$ for the YBCO $B2$
film. The bold squares represent the Hardy {\it et al.} \protect \cite{hardy}
data points on a single crystal (shifted to our $\lambda(0)$ value).
The solid line is a comparison with a $d$-wave like, strong coupling
calculation, with $2 {\it\Delta}_0=7\,k_B T_c$ (see text).
\label{HC65lambda}}
\end{figure}
\begin{table}
\caption{Characteristics of the $\rm YBa_{2}Cu_{3}O_{7}$ thin films. $d$
is their thickness and $\Delta\theta$ the rocking curve width.
$R_S$(77~K, 10~GHz) has been corrected for the film thickness.
\label{table1}}
\begin{tabular}{cccddd}
Sample&$d$ [\AA]&$T_c$ [K]&$\Delta\theta$ [$^{\circ}$]&$R_S$
[${\rm m}\Omega$]&${\sigma_{110}}$ [$\mu\Omega$cm]\\
\tableline
$A1$&1200&86&0.8&$\sim$ 10&130\\
$B1$&1400&88&0.42&0.45&\\
$B2$&1000&92&&0.27&80\\
\end{tabular}
\end{table}

\begin{references}
\bibitem{footnote} One possible origin for the extra transmission T$_0$ could
be holes in the film: indeed observations with the optical microscope show in
film $A1$ few holes of 100~$\mu$m typical size which may give rise to excess
transmission.
\end{references}
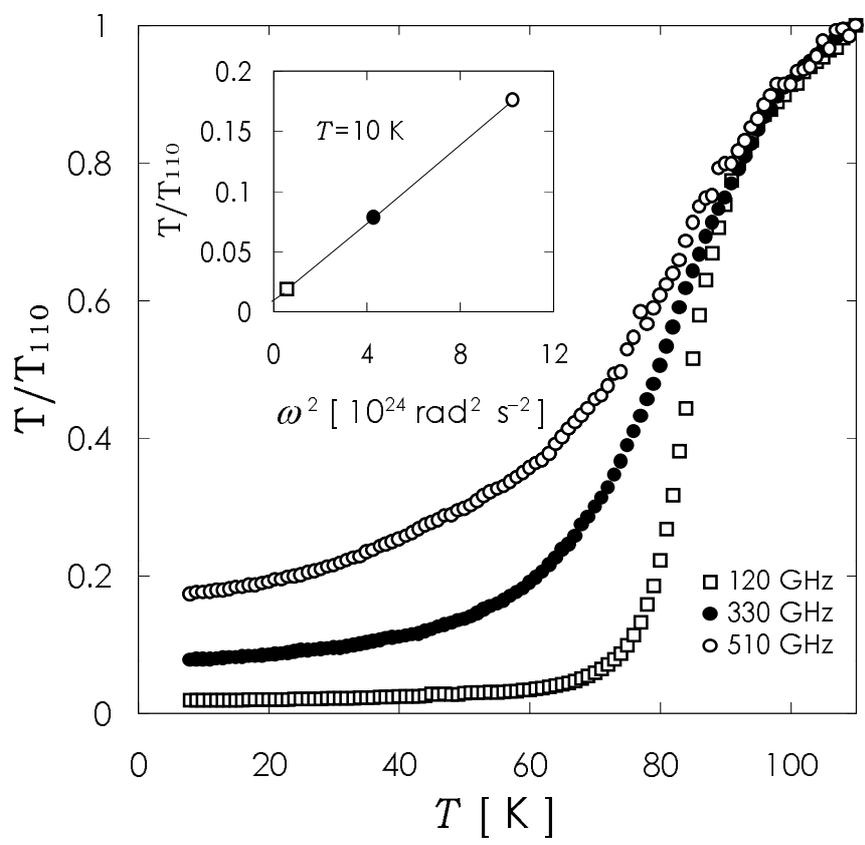
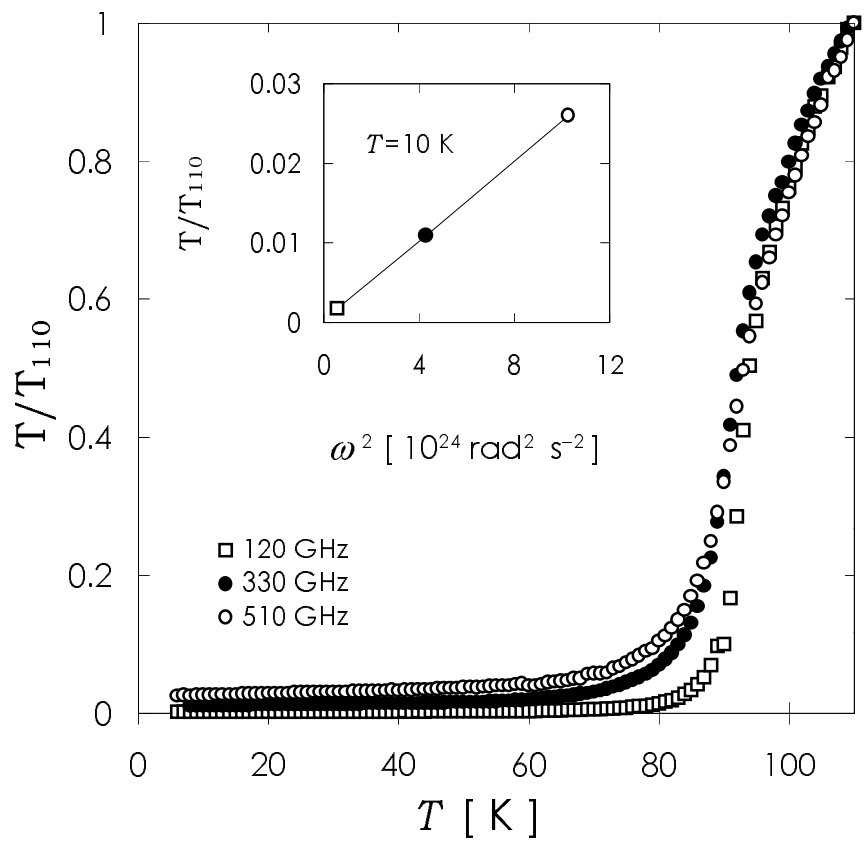
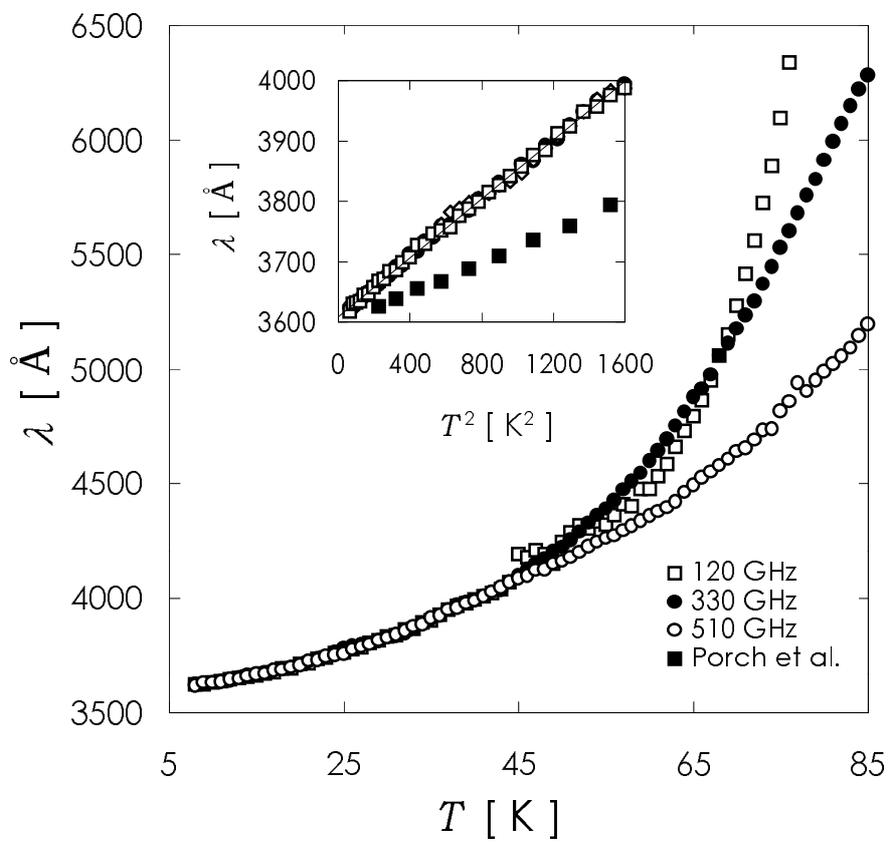
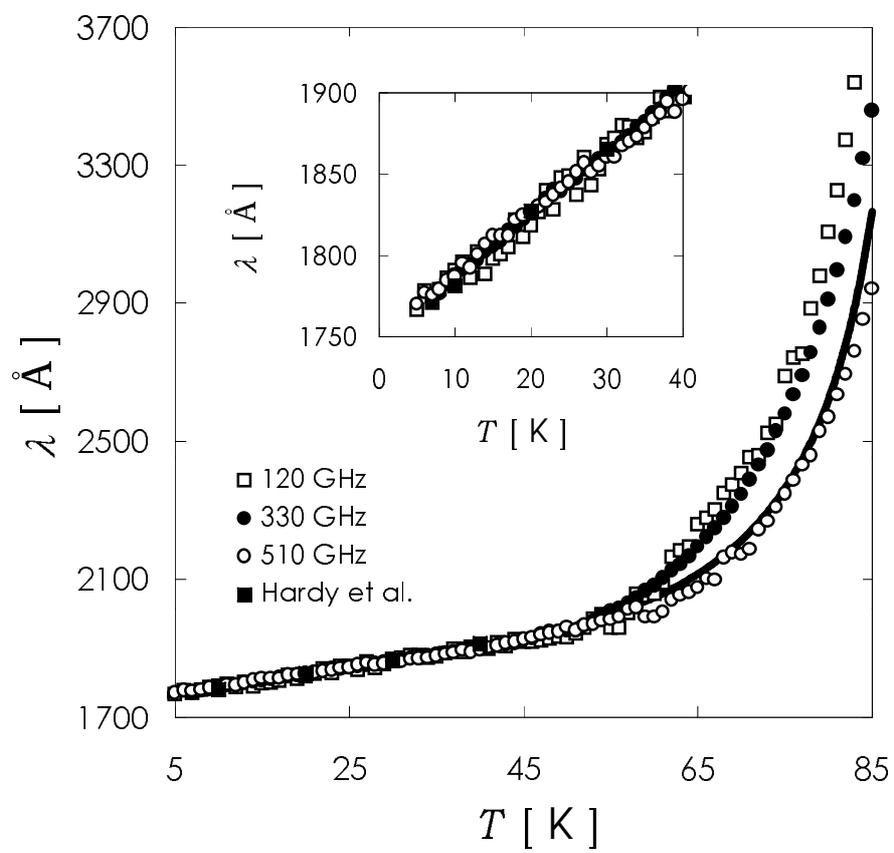
\end{document}